\documentclass[preprint]{aastex}

\newcommand{\beq}{\begin{equation}}
\newcommand{\eeq}{\end{equation}}
\newcommand{\beqn}{\begin{eqnarray}}
\newcommand{\eeqn}{\end{eqnarray}}
\newcommand{\thj}[6]{
                 \left(\begin{array}{ccc}#1&#2&#3\\#4&#5&#6\end{array}\right)
                    }

\begin{document}


\title{Statistical Power, the Bispectrum and the Search for Non-Gaussianity
                in the CMB Anisotropy}
\author{Nicholas G. Phillips}
\affil{Raytheon ITSS, Laboratory for Astronomy and Solar Physics, 
	Code 685, NASA/GSFC, Greenbelt, Maryland 20771}
\and
\author{A. Kogut}
\affil{Laboratory for Astronomy and Solar Physics, 
	Code 685, NASA/GSFC, Greenbelt, Maryland 20771}

\begin{abstract}
We use simulated maps of the cosmic microwave background anisotropy
to quantify the ability of different statistical tests
to discriminate between Gaussian and non-Gaussian models.
Despite the central limit theorem on large angular scales,
both the genus and extrema correlation
are able to discriminate between 
Gaussian models
and a semi-analytic texture model selected as a
physically motivated non-Gaussian model.
When run on the COBE 4-year CMB maps,
both tests prefer the Gaussian model.
Although the bispectrum has comparable statistical power
when computed on the full sky,
once a Galactic cut is imposed on the data
the bispectrum loses the ability to discriminate between models.
Off-diagonal elements of the bispectrum
are comparable to the diagonal elements
for the non-Gaussian texture model
and must be included to obtain maximum statistical power.
\end{abstract}



\section{Introduction}

The statistics of the cosmic microwave background (CMB)  probe
physical conditions in the early universe.  
Most inflationary models predict the CMB to follow
Gaussian statistics 
\cite{BardeenEtAl83,GuthPi82,Hawking82,Starobinski82}.
Searching CMB data for non-Gaussian distributions is thus
an important test of inflationary cosmologies.
Recent analyses of the full-sky COBE data provide mixed results.  
Kogut {\it et. al.} 
\cite{Kogut95,Kogut96} use a likelihood analysis to compare the COBE-DMR
2- and 4-year maps to the standard Gaussian model and a set of non-Gaussian toy
models.  The COBE data lie near the mode of the statistical
distributions of the genus, 3-point correlation, and correlation of extrema
points for the Gaussian model, while rejecting several non-Gaussian
alternatives. Other authors using similar statistics on the COBE maps find
comparable results 
\cite{HinshawEtAl94,Smoot94,Luo94,TorresEtAl95,ColleyEtAl96,Heavens98}.
More recently,  several authors have
challenged the conclusion that the COBE data are well described by Gaussian
statistics.  Tests using the normalized bispectrum 
\cite{FMG98,Magueijo00}
and a wavelet
analysis \cite{PVF98} both show the COBE data to be incompatible with Gaussian
simulations at 98\% confidence;
however,
these analyses did not test any non-Gaussian alternatives.
The recent work of Mukherjee (2000) brings into question the
the statistical significance of the \cite{PVF98} result.
In \cite{Mukherjee} a proper Monte-Carlo  estimation of the significance 
level shows the wavelet technique does not yield strong evidence
for non-Gaussianity in the 4-year COBE data.
Bromley and Tegmark (1999) 
argue that the
non-Gaussian signal (if any) must lie in the {\em phase} information in the
COBE sky maps, while Banday {\it et al.} (2000) 
ascribe the
discrepancy to instrumental artifacts in one of the DMR frequency  channels.

Searches for non-Gaussianity are complicated by several factors. 
On large angular scales, the superposition of multiple sources
within the instrument angular resolution 
tends to push initially non-Gaussian distributions 
toward the Gaussian model
(the Central Limit Theorem).
Sample variance can occasionally produce a non-Gaussian distribution
drawn from a Gaussian parent population.
Given enough independent tests of the same observable sky,
some outliers are expected 
\cite{BT99}. 
{\it Demonstrating that a statistic calculated for a CMB data set
lying far from the mode expected for a Gaussian sky
is a necessary but not sufficient condition
to conclude that the data are not, in fact,
a sample drawn from a Gaussian parent population.}

The statistical power of a test
quantifies the probability of correctly 
accepting or rejecting hypotheses
based on the results of that test.
Quantifying the power for a given statistic
thus requires at least two competing hypotheses.
Since there is a single Gaussian distribution
but an infinite number of non-Gaussian distributions,
we must necessarily restrict the scope of specific non-Gaussian hypotheses.
\cite{Kogut95,Kogut96} used non-Gaussian toy models
to show that the genus and extrema correlation
could discriminate between Gaussian and non-Gaussian populations.
We extend this analysis to include the bispectrum,
replacing the earlier toy models with
a model of topological defects as a specific example
of a physically-motivated alternative to the standard Gaussian model.

\section{Likelihood Analysis and Statistical Power}

Maximum-likelihood techniques are a commonly used tool 
for statistical inference. Given any 
set of $K$ statistics derivable from a sky map, 
which we view as the vector ${\bf S}$, 
we define
\beq
\chi^2\left({\bf S}\right)  = \sum_{\alpha,\beta=1}^K 
      (S_\alpha -\left<S_\alpha\right>) ({\bf M}^{-1})_{\alpha\beta}
      (S_\beta  -\left<S_\beta \right>)
\eeq
where $\left<{\bf S}\right>$ is the mean value of the statistic for the model
under consideration, and ${\bf M}$ its covariance matrix. 
The likelihood is then
\beq
{\cal L}\left({\bf S}\right) = 
   \frac{ \exp(-\frac{1}{2}\chi^2 )}
        { \sqrt{ (2\pi)^K\,\det({\bf M})}}
\eeq
where we make the convenient assumption that the 
statistics ${\bf S}$ obey a multivariate Gaussian 
probability  distribution. 
Since the large-scale CMB anisotropy is close to a Gaussian
distribution,
We do not expect the
calculation of the true distributions based on
Monte-Carlo simulations can significantly change
the relative statistical powers we determine.
For the simple comparison of two models,
we compute the likelihood functions 
${\cal L}_1$ and ${\cal L}_2$ 
for a given statistic ${\bf S}({\bf \Delta T})$. 
The data $\Delta T$ are more likely
to be drawn from parent population model 1 if 
${\cal L}_1({\bf S}(\Delta T)) > {\cal L}_2({\bf S}(\Delta T))$,
and vice versa for model 2.

For a statistic to be useful,
we need to determine how well it can discriminate between the models
under consideration.
Recall there are two types of error: 
a Type I error rejects a hypothesis when it is correct,
while a Type II error accepts a hypothesis when it is false.
The statistical power of a test quantifies the robustness of a test
against Type II errors.
By sampling the likelihood functions  
${\cal L}_i$ 
with independent samples
drawn from each of the two models under consideration,
we obtain the probability
to incorrectly identify a sample from model $i$ 
as drawn from model $j$: 
$P({\bf S}_i|H_j)$,
where ${\bf S}_i$ is extracted from a realization of model $i$.
$P({\bf S}_1|H_1)$ is thus the probability 
to correctly identify a sample from model 1 
under the hypothesis $H_1$ that it is from model 1, while 
$P({\bf S}_2|H_1)$ is the probability 
to incorrectly identify a sample from model 2
under the same hypothesis.
The probability to obtain the correct conclusion is thus
\beqn
\Pi_i &=& \frac{P({\bf S}_i|H_i)}{P({\bf S}_1|H_i) + P({\bf S}_2|H_i)} .
\eeqn
If the likelihood analysis of a CMB data set 
yields the conclusion that the data are drawn from model 1
and $\Pi_i \sim 1$, we can be confident we have made the correct
conclusion. If $\Pi_i \sim \frac{1}{2}$ then no conclusions can be made
concerning the result since we only have a 50-50 chance of being right.

\section{Models}

Any maximum likelihood analysis assumes the comparison of at least two
competing models. We use three statistical tests
(genus, extrema correlation, and bispectrum)
of the COBE-DMR 4-year sky maps 
to compare a Gaussian model
against a semi-analytic texture model,
a physical non-Gaussian theory.
We generate 1500 simulated sky maps for each model, 
and use these maps to generate the mean values $\left<S_\alpha\right> $
and the covariance matrices 
\beq
M_{\alpha\beta} = \frac{1}{N}\sum_i
       (S^{(i)}_\alpha -\left<S_\alpha\right>)
       (S^{(i)}_\beta  -\left<S_\beta \right>) ,
\eeq
including effects of instrument noise and Galaxy cut.
We then generate an additional 1500 simulated sky maps
of each of the models
for a maximum-likelihood comparison,
which we use to quantify the power of each statistical test.

The microwave anisotropy may conveniently be expressed as a sum of spherical
harmonics,
$\Delta T_i = \sum_{lm} a_{lm}\, Y_{lm}(\theta_i,\phi_i)$.
For the Gaussian model, we randomly distribute 
the spherical harmonic coefficients as
$a_{lm} = r\,\exp(i \delta)$, 
where
$\delta$ is uniformly distributed between
0 and $2\pi$ and $r$ a Gaussian random variable with zero mean and variance
\beq
\left< r^2 \right> = 
   \frac{4\pi}{5}\,\left(Q_{\rm rms}\right)^2
   \frac{
     \Gamma\left(l+(n-1)/2\right) \Gamma\left((9-n)/2\right)
        }{
     \Gamma\left(l+(5-n)/2\right) \Gamma\left((3+n)/2\right) 
        } .
\eeq
We use the best-fit spectral index $n=1.2$ 
from the COBE 4-year data
\cite{Bennett96,Gorski96,Hinshaw96,Wright96}.
Our results are insensitive to changes of $\delta n \pm 0.2$, roughly the
uncertainty in the fitted index.

Our semi-analytic model is based on probability distributions for the  texture
energy density.  We use numerical simulations of the texture field equations to
derive these distributions and in turn use these distributions to generate
simulated CMB anisotropy sky maps. The main addition to previous semi-analytic
models \cite{Turok90} is the inclusion of partial wound events \cite{Borril94}.
In \cite{PhK95}, it was verified that these events, though less energetic than
the unwinding events considered in \cite{Turok90}, are more numerous. When they
are included, the number of free parameters of the texture model increases from
just the symmetry breaking scale to now include the minimum and maximum
strengths of the partial wound events. As we discuss in the Appendix we take
the maximum strength to be the same as the unwinding strength, since these are
topological configurations that just barely become knots. The minimum strength
is put at $1/4$ the maximum.  
Our results are insensitive to the precise value chosen.

We smooth each realization of a theory sky map
with a Gaussian profile of 7\arcdeg ~FWHM to account for the
COBE-DMR beam profile. 
We add a pixel-wise noise to each map, 
$T^{\rm instr}_i = A\,T_i + n_i$, 
using the noise template appropriate for 
the 4-year COBE data as corrected for Galactic emission using
either the ``correlation'' or ``linear combination'' models
of \cite{Hinshaw96b}.
All statistics are evaluated only for the high-latitude sky
(3881 pixels at $|b| > 20\arcdeg$
with custom cutouts near Ophiuchus and Orion,
\cite{Bennett96,Banday96})
from which a fitted dipole and quadrupole have been removed.

\section{Genus}

The statistical properties of the CMB can be characterized
by the excursion regions enclosed by isotemperature contours.
The genus is the total curvature of the contours 
at fixed temperature, and may loosely be defined as the
number of isolated high temperature regions (hot spots) minus the
number of isolated low temperature regions (cold spots)
\cite{Gott90}.
The genus per unit area is a locally invariant quantity
and is insensitive to incomplete sky coverage 
(e.g. removal of the Galactic plane).

We use a nearest-neighbor search to count the number of isolated hot and cold
spots separated by temperature threshold $\nu$.  
For each map, we compute the genus at 31 values from $\nu = -3\sigma$
to $\nu = +3\sigma$, where
$\sigma = \sqrt{\left<\Delta T^2\right>}$
is the standard deviation of the noisy map.
These values form the vector  ${\bf S}$ in the likelihood analysis. 
We adopt the difference in spot number as an estimate of the genus
and use the Monte Carlo simulations to calibrate any difference 
induced by the Galactic cut
between this definition and the total curvature.
The two definitions are identical in the absence of a Galactic cut,
and the nearest-neighbor algorithm allows significant computational savings.

We use the genus to do a maximum likelihood fit for
the model amplitude for the COBE data.  We find
that both models yield similar amplitudes,
$A_{\rm fit,Gauss}=0.64 ~(Q_{\rm rms} = 13.0 ~\mu K$) and 
$A_{\rm fit,texture}=0.62 ~(G\eta^2 = 3.5\times 10^{-7}$).
The likelihood of the COBE data is greater for the Gaussian model.  
To derive the statistical power for
the genus, we present the likelihood functions with 
1500 independent sampling maps, 
with amplitudes ranging from $A=0$ to $A=3$ in 31 uniform steps. 
We fit each sample for the amplitude, 
then select the more likely model.
We then filter these samples to have  
an amplitude close to the fitted DMR map:
$|A_{\rm fit}-A_{\rm fit,Gauss}| \le 0.05$
(The filter width corresponds to our sampling interval in $A$;
our results for $\Pi_i$ are insensitive to the exact value).
We compare the number of input
models identified as each model
(Table \ref{table-genusmodel}). 
We obtain
$\Pi_{\rm Gauss} = 990/(990+704) = 58\%$.
Although the Gaussian model is
the more likely model for the COBE data, we can
not be overconfident about this result.  
When the noisier ``combination'' map is used, 
the statistical power drops to $\Pi_{\rm Gauss} = 51\%$ and  $\Pi_{\rm Texture}
= 54\%$.
Table \ref{table-statpower} shows the statistical power
for the various statistical tests.


\section{Extrema Correlation}

The extrema (peaks and valleys) of a field are defined as those points with
$\nabla T=0$. For a pixelized map, this reduces to the collection of  pixels
hotter or colder than all of their nearest neighbors. Identifying pixels hotter
than their neighbors produces a set of ``peaks'', while the locally coldest
pixels produce the ``valleys''.

The number density of peaks or valleys is dominated by the noise properties of
the map \cite{Kogut95}. The clustering of the extrema, as measured by the
2-point correlation function of the maxima and minima, provides additional
information on the underlying CMB temperature field. We define the extrema
correlation function as the 2-point function of all extrema points,
$$
C_{\rm ext}(\theta) ~=
~{ {\sum_{i,j} w_i w_j T_i T_j} \over {\sum_{i,j} w_i w_j} }
$$
(Kogut et al.\ 1995), where the sum over pixel temperatures is restricted to
the set of maxima and minima pixels separated by $\theta$
evaluated on the cut sky.
Since the correlation properties of the non-uniform noise in the DMR maps are
different from the underlying CMB temperature field, we smooth the maps with 
an additional 7\arcdeg ~FWHM Gaussian 
prior to collating the extrema as a compromise between
suppressing noise and removing CMB power at small scales,
 as was pointed out in \cite{Kogut95}.

The clustering of extrema depends on the threshold $\nu$, as defined for the
genus. We evaluate $C_{\rm ext}(\theta)$ at thresholds  $|\nu|$ = 0, 1, and 2.
Since, by definition, two peaks can not be adjacent, we ignore both the bin at
zero angular separation and the first non-zero bin in all subsequent analysis.
Simulations show that the results are dominated by the  first few remaining
bins; consequently, we speed processing by truncating the correlation function
at separation $\theta = 60\arcdeg$ 
for a total of $20$ angular bins of $3\arcdeg$ per bin
at each of the three thresholds.  We concatenate the three binned extrema
correction functions together to form our vector ${\bf S}$ for the likelihood
analysis.
A maximum-likelihood analysis of the COBE 4-year data shows the Gaussian
model to be modestly more likely than the texture model.
The statistical power of the extrema correlation is comparable to
the genus (Table \ref{table-statpower}).

\section{Bispectrum}

A third way to characterize a random field is its hierarchy of $n$ point
functions.  Since a Gaussian random field is completely
characterized  by its one and two point functions, 
the three point function or its harmonic conjugate, the bispectrum,
may prove sensitive to non-Gaussian features in the data.
We assume that the three point function only depends 
on the angular separation between points, 
\beq
C^{(3)}(\theta_{12},\theta_{13},\theta_{23}) = 
 \left<\Delta T({\bf n}_1)\,\Delta T({\bf n}_2)\,\Delta T({\bf n}_3)\right>,
\eeq
where
$\cos\theta_{ij} = {\bf n}_i\cdot{\bf n}_j$.
The expansion  in terms of Legendre polynomials \citep{Gangui94} is given by
\beqn
C^{(3)}(\theta_{12},\theta_{13},\theta_{23}) &=&  
  \sum_{l_1,l_2,l_3} \sum_{m_1,m_2,m_3} 
  \frac{2l_1+1}{4\pi}\frac{2l_2+1}{4\pi}\frac{2l_3+1}{4\pi}
   B_{l_1\, l_2\, l_3}^{m_1 m_2 m_3}\,
\\\nonumber
&&\times
  \sum_{j,k,l} \sum_{m_j,m_k,m_l} 
    P_{j}(\theta_{12})\, P_{k}(\theta_{13})P_{l}(\theta_{23})
   {\cal H}_{j\, l\, l_1}^{m_j m_l m_1}
   {\cal H}_{k\, j\, l_2}^{m_k m_j m_2}
   {\cal H}_{k\, l\, l_3}^{m_k m_l m_3}
\eeqn
with
\beq
{\cal H}_{l_1\, l_2\, l_3}^{m_1 m_2 m_3} = 
\int Y^*_{l_1 m_1}({\bf n})\,Y_{l_2 m_2}({\bf n})\,Y_{l_3 m_3}({\bf n})\ .
                                    d{\bf n}
\eeq
The bispectrum expansion coefficients are defined
in terms of the  spherical expansion coefficients $a_{lm}$ via
\beq
B_{l_1\, l_2\, l_3}^{m_1 m_2 m_3} = a_{l_1 m_1} a_{l_2 m_2} a_{l_3 m_3}.
\eeq

Our assumption that $C^{(3)}$ depends only 
on the angular separation $\theta_{ij}$
implies we are working with an isotropic cosmology.  This leads us to consider
the angular averaged bispectrum coefficients
\beq
B_{l_1\,l_2\,l_3} = \sum_{m_1,m_2,m_3} 
         \thj{l_1}{l_2}{l_3}{m_1}{m_2}{m_3}  B_{l_1\, l_2\, l_3}^{m_1 m_2 m_3}
\label{eq-bispectrum-coeff}
\eeq
where $\left( \cdots \right)$ are the Wigner 3-j symbols. We also introduce the
normalized coefficients
\beqn
I^3_{l_1\,l_2\,l_3} &=&
   \alpha_{l_1\,l_2\,l_3} \frac
     {\left| B_{l_1\,l_2\,l_3} \right| }
     {\left( C_{l_1}\, C_{l_2}\, C_{l_3} \right)^\frac{1}{2} }
\\
\alpha_{l_1\,l_2\,l_3} &=&
   \frac{1}{\left(((2l_1+1)(2l_2+1)(2l_3+1)\right)^\frac{1}{2}}
   \thj{l_1}{l_2}{l_3}{0}{0}{0}^{-1}
\nonumber
\eeqn
The diagonal ($l_1 = l_2 = l_3 = l$) coefficients are the $I^3_l$
coefficients of \cite{FMG98} 
while the $J_l^3$ coefficients of
\cite{Magueijo00} correspond to $l_1=l-2$, $l_2 = l$ and $l_3=l+2$.

To compute the bispectrum coefficients for our likelihood analysis, 
we fit the spherical expansion coefficients $a_{lm}$ 
on the cut sky
up to a maximum value
$l_{\rm max}$. 
We use these best fit  $a_{lm}$ to construct the bispectrum
coefficients  (\ref{eq-bispectrum-coeff}) and form the vector 
${\bf S} = \left\{ I^3_{lll} \right\}_{l=2,4,\ldots,l_{\rm max}}$.
With this choice, our likelihood analysis uses the same statistic as
\cite{FMG98} and we find there is no statistical power: $\Pi_{I^3_l,\rm Gauss}
= \Pi_{I^3_l,\rm texture} = 50\%$.  
Ferreira {\it et. al} (1998) 
used $l_{\rm max}=18$;
we find no change in statistical power as $l_{\rm max}$ is increased to 30.
We find a similar lack of power using 
${\bf S} = \left\{ \left| B_{lll} \right| \right\}$. 
We conclude that neither the bispectrum nor the normalized bispectrum 
in this form
can differentiate between Gaussian and texture models,
even though these models have statistically identifiable differences
in the genus and extrema correlation.

There are two possible explanations for this failure: 
i) the galaxy cut is aliasing too much power; or 
ii) we are not using enough of the available information
present in the bispectrum. 
To address these possibilities, we vary both the
size of the galaxy cut and the number of the off-diagonal terms of the
bispectrum. 
For a statistical isotropic field, 
the two-point function has 
no coupling between different $l$ multipoles: 
$\left<a_{lm}\, a^*_{l'm'} \right> = C_l\,\delta_{ll'}\,\delta_{mm'}$. 
We find this assumption can not be made for
the three point function. 
Figure \ref{fig-bispec-distri}a shows the
distribution of  $\left<\left|B_{l_1\,l_2\,l_3}\right|\right>$ values for an
$n=1$ Gaussian  model without noise or a galaxy cut (averaged over 1500
realizations). The diagonal values,
$\left<\left|B_{l\,l\,l}\right|\right>$,  are dominant,  
but the off-diagonal terms also have non-zero values
comparable to the diagonal. Moreover, the values group according to how much
they are off the diagonal. There are three groups: 
the diagonal values,
the terms off just one diagonal 
($\left<\left|B_{l\,l,l'\ne l}\right|\right>$), 
and finally the terms completely  off the diagonal\footnote{
We only consider the terms that are not forced to zero 
by the triangularity of the 3-j symbols.}
($\left<\left|B_{l_1\,l_2\,l_3}\right|\right>$, $l_1\ne l_2 \ne l_3$). 
Figure \ref{fig-bispec-distri}b shows the
distribution for our texture model. 
The bispectrum coefficient values still cluster, but 
not as tightly as the Gaussian model.
The $J_l^3$ coefficients of \cite{Magueijo00} are in this
last group. Their choice of statistic misses the more significant
set of terms off just one diagonal. 
The normalized bispectrum coefficients
$\left<I^3_{l_1\,l_2\,l_3}\right>$  
demonstrate similar clustering properties.

The Galactic cut destroys the bispectrum's ability
to differentiate models. 
Figures \ref{fig-bispec-distri}c and \ref{fig-bispec-distri}d show the
distributions of $\left<\left|B_{l_1\,l_2\,l_3}\right|\right>$ for Gaussian and
texture models after a galaxy cut is included. 
The values no longer cluster,
and are no longer identifiable by model.
The spherical harmonics $Y_{lm}$ are no longer orthogonal on the cut sky;
the resulting aliasing of power between modes
overwhelms the signal.

To quantify this failure, we compute the statistical power as we impose the
simple latitudinal cut of removing pixels with $\theta \le b$, for $b$ varying
from $0\arcdeg$ to $14\arcdeg$.  We set a signal amplitude that corresponds to the genus
and extrema correlation function ($A=0.64$). 
Since the bispectrum computed using only the diagonal values
never shows statistical power, we include 
both the diagonal and once-off-the-diagonal terms 
(the two rightmost groups in Fig \ref{fig-bispec-distri}). 
Figure \ref{fig-bispectrum-galcut} shows that 
the probability to correctly identify the input model
falls monotonically as the Galactic cut is increased.
For comparison purposes, we also plot the statistical power for the genus,
which shows only a slight trend consistent with larger sample variance
as the solid angle available for analysis is reduced.
With no galactic cut, the bispectrum has comparable power 
as the genus\footnote{Since we fix the input amplitude instead of fitting,
the genus power here is slightly larger 
than shown in Table \ref{table-statpower}.}.
The statistical power of the
bispectrum drops steadily until cut $b<12\arcdeg$, where there is no power. 
The normalized bispectrum $I^3_{l\,l\,l'}$ is {\it never} able to
differentiate between the models, even when there is no galaxy cut present. 

The other  important consideration  is how much information is needed.  For
$l_{\rm max}=20$, there are 770 nonzero, unique bispectrum values, of which 10
are the non-zero diagonal values. 
Using all values instead of just the diagonal values 
is computationally expensive.
We use instead the diagonal terms 
and terms with one off-diagonal $l$ value
(second group from the right in Figure \ref{fig-bispec-distri})
to form the vector 
${\bf S} = \left\{\left|B_{l\,l\,l'}\right| \right\}$. 
The restriction $|l-l'|\le\Delta l$
selects how many of the off-diagonal terms we use. 
$\Delta l=0$ corresponds to the analysis of \cite{FMG98} 
and $\Delta l=l_{\rm max}$ includes all 
the once-off-the-diagonal terms. 
Figure \ref{fig-bispectrum-deltal} shows the
statistical power of the bispectrum as we increase $\Delta l$, 
where we consider
both noisy and noiseless data sets evaluated over the full sky.
The power is greater without noise;
the statistical power of the Gaussian model
is fairly constant after the inclusion of
just a few extra off-diagonal terms. For the texture mode, the inclusion of
additional off-diagonal terms steadily improves the probability for correct
identification until $\Delta l\sim l_{\rm max}/2$. With noise, the statistical
power for either model levels out at about the same value as the genus and
extrema correlation function. 
This suggests that 
but for the need to include a galaxy cut, 
a likelihood analysis based on $|B_{l\,l\,l'}|$ would be as
powerful as the other tests.  
The galaxy cut aliases power in the conjugate harmonic space, 
but has negligible effect in the coordinate space 
used by the genus and  extrema correlation function.

The addition of off-diagonal terms is important for the non-Gaussian
texture model.
For ideal (noiseless) data, 
the probability to correctly identify the texture model 
goes from 59\% for only
diagonal terms to 74\% when all $|B_{l\,l\,l'}|$ terms are considered. 
In contrast, the probability to correctly identify the Gaussian model
is nearly constant.
Since the three point function contains
no new information for a Gaussian theory,
we do not expect much improvement in the Gaussian model
as the number of terms is increased. 
The texture model was
chosen exactly because it is an example of a non-Gaussian theory,
so the three point function or bispectrum should be sensitive to the gradual 
addition of the information present in the off-diagonal terms.

We have duplicated all the analysis discussed above for the  normalized
bispectrum coefficients $I^3_{l\,l\,l'}$ introduced by \cite{FMG98}.  Figs
\ref{fig-bispectrum-galcut} and \ref{fig-bispectrum-deltal} show the likelihood
analysis in terms of the normalized bispectrum
never has any statistical power.  For the
ideal situation of noiseless data without a galaxy cut, the probability to
correctly identify either model is at best 52\%. 
The statistic $J_l^3$ of \cite{Magueijo00} corresponds to our
$I^3_{l-2,l,l+2}$ and is thus not directly covered in our analysis.
Figures \ref{fig-bispec-distri}a and \ref{fig-bispec-distri}b show
$J_l^3$ to be of lesser significance than our most general set
of $I^3_{l\,l\,l'}$'s and we can not expect any statistical power
for this choice.
Although we have tested only a single non-Gaussian model,
the normalized bispectrum does not appear to be a promising test
of the statistical distribution of CMB anisotropy.


\section{Discussion.}

Recent work has generated new interest in statistical tests of the CMB.
Comparison of the bispectrum of the COBE-DMR 4-year sky maps
to simulations of Gaussian maps show the COBE data to lie far from the
mode of the Gaussian models \cite{FMG98}.  
Since this work tested only Gaussian models, 
it is not clear whether the discrepant values
result from non-Gaussian signals in the data
(including instrumental artifacts)
or are simply an outlier drawn from a Gaussian parent population.
We compute the statistical power of several popular statistics
(the genus, extrema correlation, and bispectrum)
to determine their ability to distinguish between
a Gaussian model and a texture model chosen as a physically motivated
non-Gaussian alternative.

Both the genus and extrema correlation successfully discriminate
between the models with 60\% confidence,
limited by the similarity of the texture model to a Gaussian
on large angular scales (the central limit theorem).
The COBE 4-year data prefer the Gaussian model,
although the modest statistical power
prevents a strong rejection of this specific non-Gaussian alternative.
In contrast, we find that the bispectrum
has no power to discriminate between models,
and trace this to the aliasing of power
caused by the Galactic cut imposed on the data.
The normalized bispectrum {\it never} has any power,
even run on noiseless simulations with no Galactic cut.
Since the results presented in \cite{FMG98} and \cite{Magueijo00}
are based on an analysis in terms of the normalized bispectrum,
the results, though correct, have no statistical power. 
Their results provide a necessary, but not sufficient, condition
to rule out an underlying Gaussian character to the CMB anisotropy.

The problem of aliased power can in principle be solved,
either by windowing the multipole coverage
or by construction of an explicitly orthogonal basis.
With this in mind, we further explore the statistical power
of the bispectrum using full sky coverage to avoid
the problem of aliased power.
We find significant coupling between multipoles in the bispectrum coefficients
for the non-Gaussian (texture) model.
Failure to include these off-diagonal terms
reduces the statistical power of the bispectrum on the full sky
from 67\% (comparable to the genus or extrema correlation)
to 55\% (not much better than random guessing).
The off-diagonal elements are more important
for the non-Gaussian model,
pointing out the importance of 
testing various statistics against some alternative
to the standard Gaussian model.


\newpage
\appendix
\large
\begin{center}{\bf Appendix}\end{center}
\normalsize
\section{Texture Model}

In order to test the power of different statistics,
we need an alternative hypothesis to the standard Gaussian model.
Defect models
\cite{Kibble85a,Kibble85b,VS95}
are the only known cosmological alternatives
expected to provide a significant non-Gaussian signal.
Characterizing this non-Gaussian signal
requires many realizations of the model.
Unfortunately, the non-linear nature of defect models
make it impossible to carry out the analysis in closed form,
while also making exact numerical work computationally expensive.
We use exact simulations of a texture model
to derive statistical descriptions of the energy density,
then use these statistical descriptions
to rapidly generate additional realizations
of the model.

Textures \cite{Turok89} are a class of topological defects 
in which the vacuum manifold of the order field in broken symmetric phase  
is of the same dimension as the spatial geometry.
This allows the order field to remain on the vacuum manifold,
regardless of initial conditions.
Perturbations in the energy density
are driven solely by the order field's kinetic energy.
As the correlation length of the order field grows, 
causal regions emerge that posses non-trivial topological charge (knots).
These configurations collapse until the energy density is great enough
for the order field to tunnel through the vacuum manifold and unwind.  
As CMB photons travel through these evolving regions of increased energy 
density,  
they become blue/red shifted and the anisotropy arises. 
This basic picture of knots in the texture order field 
led to the first approximation of Turok and Spergel (1990) 
which randomly placed idealized knots across the sky.
The resulting CMB anisotropy, however, 
did not agree with results 
derived from full numerical simulations \cite{UPen94}.
   
Borril {\it et. al.} (1994) 
suggested the existence of
a less energetic but more numerous configuration:
partially wound events (PWEs),
in which order field wraps around the vacuum manifold
but not enough to be knotted.
In \cite{PhK95}, we directly verified, 
for an expanding flat universe,  
that PWEs are far more numerous than knots and 
that their contribution dominates the CMB anisotropy.  
In this Appendix, we provide
a statistical description of the PWEs.
We then extend the model of \cite{Turok90}
by randomly placing texture events on the microwave sky,
drawing each event from the correct distribution in energy.

\subsection{Field Equations and Numerical Implementation}

Since we are interested in times 
long after the phase transition for the order field, 
we treat the field as classical. 
Except for the special case where the field unwinds, 
we restrict the field to lie on the vacuum manifold. 
We will focus solely on textures, 
so the (real-valued) global order becomes
$\tilde{\bf \Phi} = \left(
  \tilde\Phi_1,\tilde\Phi_2,\tilde\Phi_3,\tilde\Phi_4\right)$ and the vacuum
manifold is the three sphere $S^3$: 
$|\tilde{\bf\Phi}|^2=\phi_0$. 
After re-scaling the field
${\bf\Phi} = \tilde{\bf\Phi}/\phi_0$,
the field dynamics are determined by the action
\begin{equation}
S[{\bf\Phi}] = \int d^4x\sqrt{-g}
\left[
  \frac{1}{2}\left(\partial_\mu {\bf\Phi}\right) \!\cdot\!
             \left(\partial^\mu {\bf\Phi}\right)
  -\lambda\left(|{\bf\Phi}|^2 - 1\right)
\right].
\end{equation}
The constraint that the field stay on the vacuum manifold is imposed via the 
Lagrange multiplier $\lambda$. This is the nonlinear $\sigma$ model (NLSM)
(see e.g. \cite{UPen94}). 
We carry out our work for a spatial flat Robertson-Walker
homogeneous universe.
For this choice the field equations become
\begin{equation}
\ddot{\bf\Phi} + 2\frac{\dot a}{a}\dot{\bf\Phi} - \nabla^2{\bf\Phi}
= \left( \left|\nabla{\bf\Phi}\right|^2 - \left|\dot{\bf\Phi}\right|^2
    \right){\bf\Phi}
\label{eq-feqn}
\end{equation}
where $\dot{\bf\Phi} = \partial{\bf\Phi}/\partial\tau$ and 
$\nabla{\bf\Phi} = \partial{\bf\Phi}/\partial x_i$ is the
spatial gradient. We refer the reader to \cite{PhK95} for details 
of the derivation of discrete versions of these equations
and their numerical implementation.

We characterize the field dynamics using two quantities.
The first is the energy density, the $\tau\tau$ component of the
field stress tensor:
\begin{equation}
\rho_{i,n} = \frac{1}{2}\left(
   \left|\dot{\bf\Phi}_{i,n}\right|^2 + \left|\nabla{\bf\Phi}_{i,n}\right|^2
\right)
\end{equation}
with the obvious discrete versions of the derivatives,
where index $i$ labels the spatial grid cells and $t$ is the temporal label.
The second quantity is the cell alignment,
\begin{equation}
\alpha_i = \left( \frac{1}{6} \sum_j {\bf\Phi}_j\right) \!\cdot\! {\bf\Phi}_i
\end{equation}
where $j$ runs over cell $i$'s six nearest neighbors. 
With this definition, 
the alignment of a grid cell is the cosine of the angle between the field
configuration at $i$ and the average of the field over $i$'s nearest neighbors.
$\alpha_i$ measures how close the vacuum at $i$ is to its neighbors, or how
aligned the cell is with the local choice of vacuum. This gives a rough
measure of the local topological charge 
(see \cite{Borril94b} for further discussion). 
According to this use of the NLSM
model, knots are numerically identified by $\alpha_i < 0$  \cite{UPen94}.
They are unwound by flipping the field  to its anti-podal: 
${\bf\Phi}_i\rightarrow -{\bf\Phi}_i$. 

We base our analysis  on a catalogue of events identified in 
4000 simulations of the order field dynamics. 
The expansion scale factor $a(t)$ corresponds to a
matter-dominated universe.
We run the simulations on a $64^3$ grid array with 
periodic boundary conditions, 
continuing until the horizon crossing time ($\Delta\tau=32$).
The initial order field configuration 
$\Phi({\bf x},\tau_0)$ is uniformly distributed over the vacuum manifold. 
We begin the simulation by integrating this initial random configuration,
but with the anti-podal flipping of knot configurations disabled. 
This acts to ``smooth'' the initial order field 
while maintaining the constraint $|{\bf\Phi}|^2 = 1$. 
We then re-enable the anti-podal unwinding and
begin the event identification procedure outlined below.  


\subsection{Event Identification}

As the
casual horizon grows and regions 
of different vacua come
into contact, the order field ${\bf\Phi}({\bf x})$  changes
until it is in the same vacuum throughout. As this
takes place  the  energy density due to the field rises. The dynamics of
${\bf\Phi}$ force the region 
containing the changing ${\bf\Phi}$ to
shrink in size  \cite{Derrick64}. 
When the total
topological charge contained in this region is non-trivial, 
i.e., a knot,
the energy density
within the shrinking region will increase
until the field can tunnel through the vacuum
manifold and unwind the topological charge.
The other possibility is that the order field within the region,
though needing to 
resolve which vacuum the region will occupy, 
has no net topological charge.
These are the partially wound events. 
They too are shrinking regions of growing energy,
but now order field sorts itself out 
without having to tunnel through the vacuum manifold. 
In either case, after the region shrinks and resolves which vacuum to
occupy, the region  then  expands as the stress energy diminishes.

We use this behavior to 
identify events in the numerical simulations.
At each time step, 
we record the position, alignment, and energy density
for each local maximum in the energy density, 
then track the spatial movement of each local maximum from
one time step to the next.
These local maxima are the grid centers of the growing/shrinking 
regions discussed above. 
We identify 300597 events in the 4000 simulations run.
Figure \ref{fig-sample-events} shows
the temporal evolution of typical events.
The alignment decreases and the energy increases
as the texture becomes increasingly wound.
Shortly after the alignment reaches a minimum,
the energy density begins to decrease and the event dissipates.

Fig.\ \ref{fig-total-align} shows the distribution of these events
sorted by the minimum alignment $\alpha_{\rm min}$ of each event. 
Three groupings are evident: 
knots, PWEs and noise. 
The knots are all events with minimum alignment less than zero.  
We identify all events with $\alpha_{\rm min} > 0.55$ as noise,
since they have negligible contribution to the energy of the field.
We identify all remaining events,
$0 < \alpha_{\rm min} \le 0.55$, as partially-wound events.
With this identification, 4\% of the events are knots, 
56\% are PWEs, and 40\% are noise.


\subsection{Partially Wound Event Distributions.}


A semi-analytic model for textures requires
a statistical description of the temporal and energy distribution 
for the PWEs. 
Instead of working directly with the energy distribution, 
we consider the alignment of the PWEs.
The alignment is more closely related to the topological properties of 
the theory and as such is less sensitive to 
noise in the simulated order field.

We label each event by the minimum of its alignment, $\alpha_{\rm min}$, 
and the conformal time $\tau_{\rm min}$ of this minimum alignment. 
Binning the PWEs according to $\alpha_{\rm min}$ and $\tau_{\rm min}$
shows the distribution to be separable, 
$dN_{\rm PWE}(\alpha,\tau) \propto dN_1(\tau)\, dN_2(\alpha)$.
Not surprisingly, 
the temporal distribution for the PWEs has the same behavior 
as that for the knots,
\begin{equation}
dN_1(\tau) = \frac{\kappa_{PWE}}{\tau^4}d\tau;
\quad \kappa_{\rm PWE} = \frac{3}{5}.
\label{eq-temporal-dis}
\end{equation}
We determine $\kappa_{\rm PWE}$ by comparing the total number of
PWEs ($N_{\rm PWE}=168307$) to the number knots ($N_{\rm knot}=11295$).
There are 15 partially wound events for each unwinding event.

Figure \ref{fig-alignment} shows the distribution
in alignment, $dN_2(\alpha)$.
The shape is well described by a $\chi^2$ distribution
with 4 degrees of freedom,
\begin{equation}
dN_2\left(\alpha\right) = A \left[ \chi^2_\nu(x) + p_0 \right]
\label{eq-align-dis}
\end{equation}
where
$ \chi^2_\nu(x)$ is the $\chi^2$ distribution for $\nu=4$ degrees of freedom,
\begin{equation}
x = \frac{{\bar\alpha}-\alpha}{m},
\end{equation}
${\bar\alpha} = 0.55$ is the alignment noise cutoff,
and $A$ is a normalization constant.
The uncertainty in the number of events $N_j$ in each bin
is given by counting statistics,
$\sigma^2_{N_j} = N_j$.
We thus fit the binned data to determine values
\begin{equation}
m = 0.165\pm 0.003
\quad{\rm and}\quad
p_0 = 0.026\pm 0.005
\end{equation}
A four-parameter fit with $\nu$ and ${\bar\alpha}$ free
yields fitted values
$\nu = 4.01 \pm 0.02$ and ${\bar\alpha} = 0.54 \pm 0.01$,
justifying the choice of functional form 
of the probability distribution.

We also test the spatial properties of the partially-wound events,
to determine whether one PWE will suppress or enhance the probability 
to find another PWE in the local neighborhood.
The two-point correlation function for all PWEs 
is indistinguishable from one derived
for the same number of events uniformly distributed in space.
A principal moment analysis of the spatial distribution of the energy
density around the gird center shows the PWEs 
to be spherically symmetric.

\subsection{Sky Map Generation}

Our semi-analytic model uses this statistical description of the order field
in texture cosmologies to generate maps of simulated CMB anisotropy
without requiring computationally expensive numerical simulation 
of the evolving order field.
It extends the work of Turok and Spergel (1990) 
to include the more numerous partially-wound events.
As in that work, we generate a random set of 
of spherically symmetric events 
with uniform spatial distribution and $\tau^{-4}$ temporal distribution.
Now, however,
the energy of each event is drawn either
from a mono-energetic knot population (6\% probability)
or
from the distribution of PWEs (94\% probability).

Deriving a map of the CMB anisotropy from a realization
of random texture events
requires fixing the relation between the minimum alignment $\alpha$
used to generate the PWE distribution
(Eq. \ref{eq-align-dis})
and the amplitude $\Delta T/ T$ of the CMB anisotropy.
Analysis of the simulations indicates this 
relation to be linear.
Due to the discrete anti-nodal flipping algorithm,
the energy scale of the {\it knots} in the simulations is arbitrary.
We thus require two normalization parameters to fully specify
CMB anisotropy given a distribution of texture events:
the minimum and maximum energy of the PWEs
relative to the knots.
We take the maximum amplitude for PWEs
to be equal to the amplitude for the knots:
these represent (rare) events that just barely missed
unwinding ($\alpha\sim 0$). 
The noise cutoff $\alpha_{\rm min} = 0.55$
provides a lower limit to the energy of PWEs used in the model;
we thus take the minimum amplitude 
to be half way between the knot amplitude and zero.

A single realization of a CMB map using the semi-analytic model
thus requires the following steps:
\newline\indent
1) we compute the number of events in the observed universe by taking the
product of the observed volume and the integrated temporal distribution
(\ref{eq-temporal-dis}); 
\newline\indent
2) we distribute these events uniformly in space
and with a $\tau^{-4}$ temporal distribution; 
\newline\indent
3) letting 6\% of these be knots,
we assign them the maximum amplitude and let the other 94\% be PWEs. For
the PWEs, we select random alignments from the distribution 
(\ref{eq-align-dis})
and linearly map to amplitudes as outlined above; 
\newline\indent
4) we use the result from \cite{Turok90} for the $\Delta T/T$ 
contribution to determine which pixels are influenced by
each event.
\newline
We are thus
left with one undetermined parameter, the amplitude of
the knotted configurations.
A single full-sky realization of the semi-analytic model
with 6144 pixels
requires only 20 seconds of CPU time
on a Sun Ultra workstation
at 360 MHz clock speed,
achieving the goal of rapidly
generating a large number of sky maps.


\acknowledgments{
We thank Ue-Li Pen for valuable discussions. 
This work was funded in part through 
NASA RTOP 399-20-61-01.
}

\newpage

\figcaption{
Distribution of average bispectrum values 
$\left<\left|B_{l_1\,l_2\,l_3}\right|\right>$ for 1500 maps.
(a) Noiseless $n=1$ Gaussian model, no galaxy cut.
(b) Noiseless $E_{\rm min}=0.25\,E_{\rm max}$ texture model, no galaxy cut.
(c) Same as (a), but now with a galaxy cut.
(d) Same as (b), but now with a galaxy cut.
(For ease of viewing, each of the bispectrum was rescaled
to a common range. All analysis was done without
this rescaling.)
\label{fig-bispec-distri}
}

\figcaption{
Dependence of the 
statistical power of the bispectrum $\left|B_{l_1\,l_2\,l_3}\right|$
and power-spectrum normalized version $I^3_{l_1\,l_2\,l_3}$ 
on the size $b$
of the galactic cut. The statistical power is measured by the probability
to correctly identify the input model.
\label{fig-bispectrum-galcut}
}

\figcaption{
Statistical power of the bispectrum $\left|B_{l\,l\,l'}\right|$ and
normalized bispectrum $I^3_{l\,l\,l'}$ as a function
of the number of off-diagonal terms.
The number of off-diagonal terms is controlled by the restriction
$|l-l'| \le \Delta l$.  $\Delta l = 0$ uses ony the diagonal terms
and is appropriate only if there is no coupling between multipole
terms in the bispectrum.
The texture model (triangles) has significant contribution
from off-diagonal terms.
The normalized bispectrum has no power regardless of the number of
off-diagonal terms.
\label{fig-bispectrum-deltal}
}

\figcaption{
Temporal evolution of typical texture events. The top panel shows
the alignment at the center of each event at each time step. 
The bottom panel shows the energy density at the center, 
defined as the local maximum of the energy density.
\label{fig-sample-events}
}

\figcaption{
Distribution of texture events sorted by the minimum alignment reached
over the history of each event.
Three peaks are evident.
Events with $\alpha_{\rm min} < 0$ correspond to knots.
The energy density at the time of minimum alignment (bottom panel)
shows that the region $\alpha_{\rm min} > 0.55$ contributes negligible.
We identify these events as noise.
The middle region corresponds to partially-wound events (PWEs).
\label{fig-total-align}
}

\figcaption{
The distribution function $dN_2(\alpha)\Delta\alpha$ for 
partially-wound events.
The smooth curve is the fitted $\chi^2_4$ distribution 
with 4 degrees of freedom.
\label{fig-alignment}
}

\newpage
\begin{deluxetable}{lcc}
\tablewidth{0pt}
\tablecaption{Distribution of Model Identifications}
\tablehead{
  \colhead{} & \multicolumn{2}{c}{Identified as} \\
  \colhead{Input Model} & \colhead{Gaussian} & \colhead{Texture}
}
\startdata
   Gaussian  & 990 & 451 \\
   Texture   & 704 & 845 
\enddata
\label{table-genusmodel}
\end{deluxetable}

\begin{deluxetable}{cccccc}
\tablewidth{0pt}
\tablecaption{Statistical Power of Analyzed Tests}
\tablecolumns{6}
\tablehead{
  \colhead{} & \colhead{} & \colhead{} & \colhead{} &
  \multicolumn{2}{c}{Probability Right} \\
  \colhead{Statistic}    & \colhead{Noise Map} & \colhead{Amplitude} &
  \colhead{Galaxy Cut} & \colhead{Gaussian}    & \colhead{Texture}
}
\startdata
  Genus     &  Corr & $A_{\rm fit}=0.64$& Template &   58     &  65     \\
  Genus     &  Comb & $A_{\rm fit}=0.66$& Template &   51     &  54     \\
 Extrema    &  Corr & $A_{\rm fit}=0.61$& Template &    61    &  66     \\
 Extrema    &  Comb & $A_{\rm fit}=0.72$& Template &    54    &  55     \\ 
\cutinhead{Bispectrum with varying Galaxy Cut}
  Genus     &  Corr & $A = 0.64$  & $b\le 14^0$ &   63     &   62    \\
$B_{lll'}$
            &  Corr & $A = 0.64$  & $b\le 14^0$ &   52     &   51    \\
$I^3_{lll'}$
            &  Corr & $A = 0.64$  & $b\le 14^0$ &   51     &   51    \\
  Genus     &  Corr & $A = 0.64$  &  None       &   68     &   67    \\
$B_{lll'}$
            &  Corr & $A = 0.64$  &  None       &   64     &   64    \\
$I^3_{lll'}$
            &  Corr & $A = 0.64$  &  None       &   51     &   51    \\
\cutinhead{Bispectrum with and without Off-Diagonal Terms, with Noise}
$B_{lll}$
          &  Corr &  $A = 0.64$   &   None     &   62     &  55     \\
$B_{lll'}$
          &  Corr &  $A = 0.64$   &   None     &   64     &  64     \\
$I^3_{lll}$
          &  Corr &  $A = 0.64$   &   None     &   50     &  50     \\
$I^3_{lll'}$
          &  Corr &  $A = 0.64$   &   None     &   51     &  51     \\
\cutinhead{Bispectrum with and without Off-Diagonal Terms, Noiseless}
$B_{lll}$
          &  None &               &   None     &   68     &  59     \\
$B_{lll'}$
          &  None &               &   None     &   71     &  74     \\
$I^3_{lll}$
          &  None &               &   None     &   50     &  50     \\
$I^3_{lll'}$
          &  None &               &   None     &   52     &  52     \\
\enddata
\label{table-statpower}
\end{deluxetable}

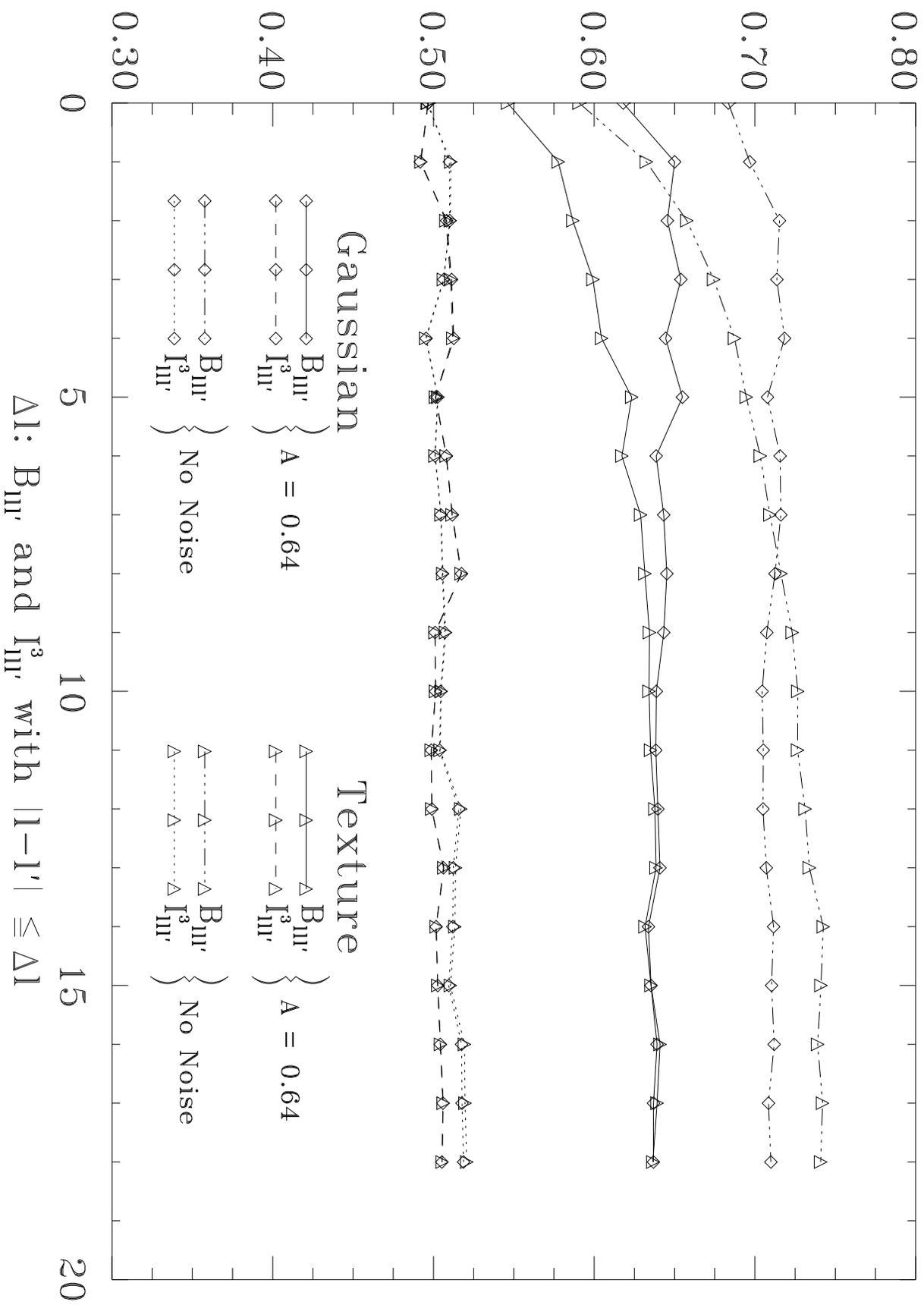
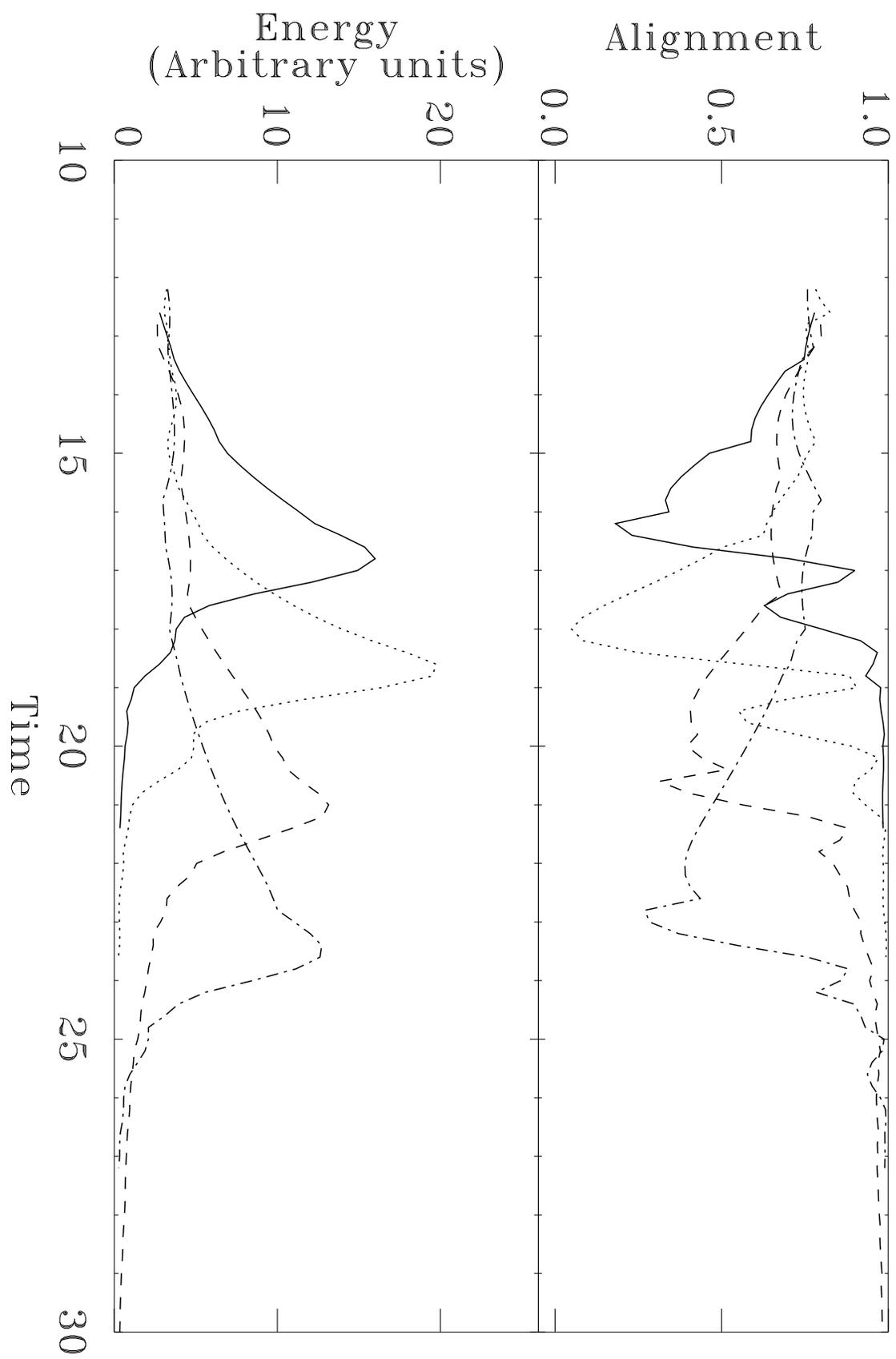
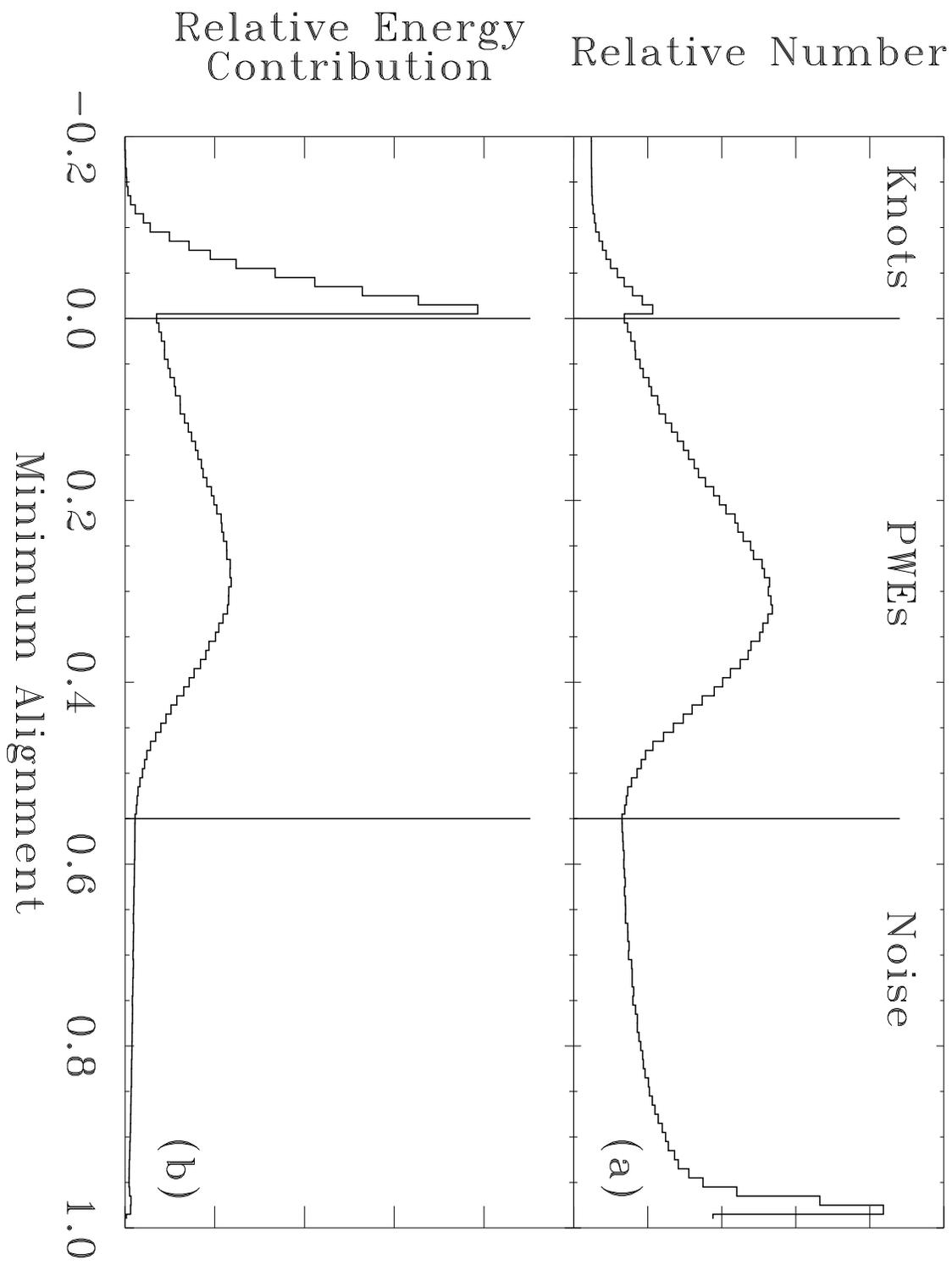
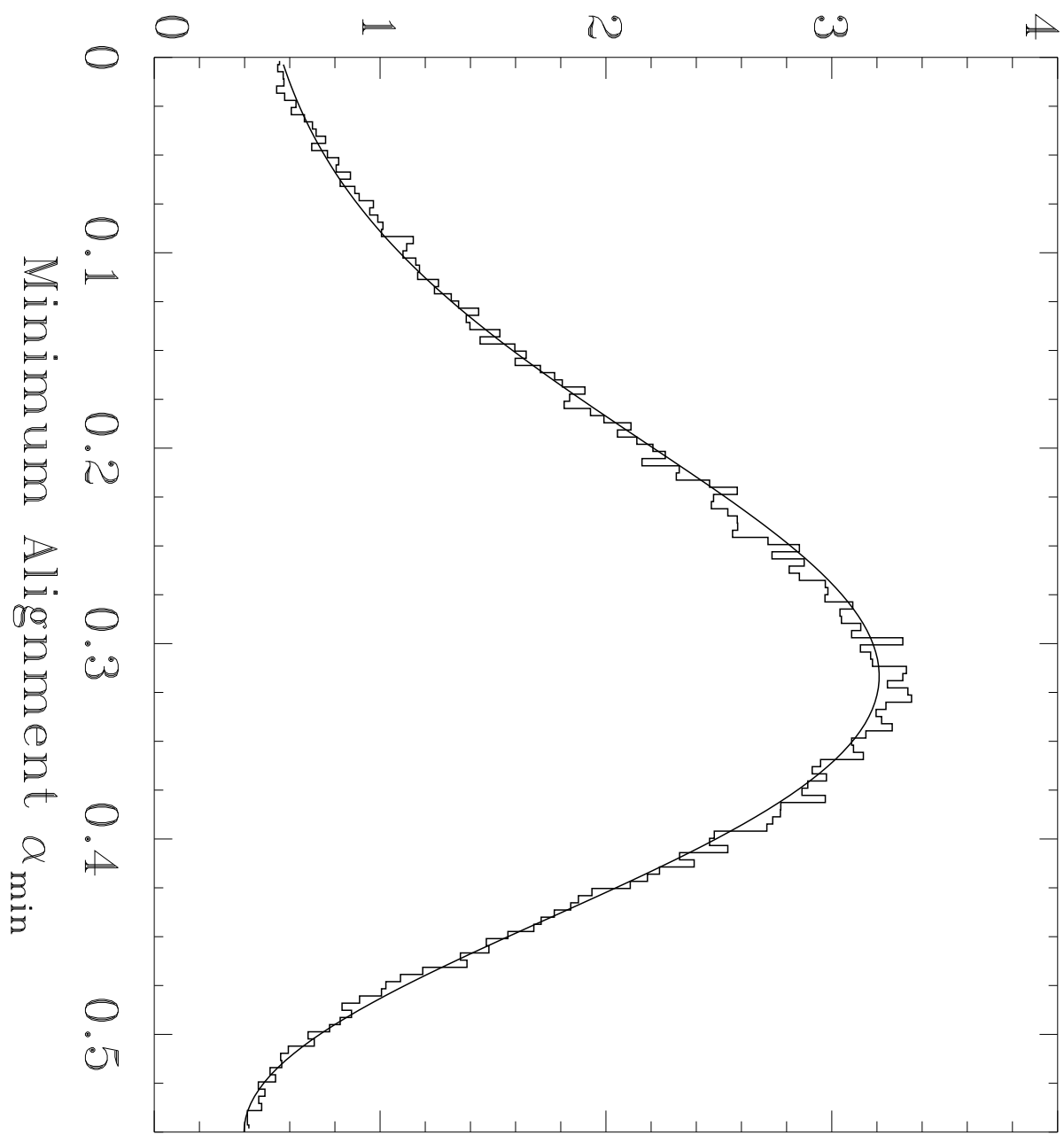
\end{document}